\documentclass[preprint,onecolumn,letterpaper, superscriptaddress, longbibliography ]{revtex4-1}

\usepackage{graphicx}
\usepackage{dcolumn}
\usepackage{bm}

\usepackage{graphicx}
\usepackage{bm}
\usepackage{amssymb}
\usepackage{amsmath}
\usepackage{dcolumn}
\usepackage{subfigure}
\usepackage{threeparttable}
\usepackage{multirow}
\usepackage[usenames,dvipsnames]{color}
\usepackage{wasysym}
\usepackage{txfonts}
\usepackage{mathtools}

\begin{document}

\preprint{ }

\title[title]{Magnon-magnon interactions induced by spin pumping-driven symmetry breaking in synthetic antiferromagnets}

\author{M. M. Subedi}
\affiliation{Department of Physics and Astronomy, Wayne State University, Detroit, MI 48202, USA
}
\author{K. Deng}
\affiliation{Department of Physics, Boston College, 140 Commonwealth Avenue, Chestnut Hill, Massachusetts 02467, USA
}

\author{Y. Xiong}
\affiliation{Department of Physics, Oakland University, Rochester MI 48309, USA}
\affiliation{Department of Electronic and Computer Engineering, Oakland University, Rochester MI 48309, USA}

\author{J. Mongeon}
\affiliation{Department of Physics, Boston College, 140 Commonwealth Avenue, Chestnut Hill, Massachusetts 02467, USA
}

\author{M. T. Hossain}
\affiliation{Department of Physics and Astronomy, University of Delaware, Newark, DE 19716, USA}

\author{P. B. Meisenheimer}
\affiliation{Department of Materials Science and Engineering, University of Michigan, Ann Arbor, MI
48109, USA}

\author{E. T. Zhou}
\affiliation{Department of Physics and Astronomy, Wayne State University, Detroit, MI 48202, USA
}

\author{J. T. Heron}
\affiliation{Department of Materials Science and Engineering, University of Michigan, Ann Arbor, MI
48109, USA}

\author{M. B. Jungfleisch}
\affiliation{Department of Physics and Astronomy, University of Delaware, Newark, DE 19716, USA}

\author{W. Zhang}
\affiliation{Department of Physics, Oakland University, Rochester MI 48309, USA}
\affiliation{Department of Physics and Astronomy, University of North Carolina, Chapel Hill, NC 27599 USA}

\author{B. Flebus}
\affiliation{Department of Physics, Boston College, 140 Commonwealth Avenue, Chestnut Hill, Massachusetts 02467, USA
}

\author{J. Sklenar}
\affiliation{Department of Physics and Astronomy, Wayne State University, Detroit, MI 48202, USA
}

\begin{abstract}
The richness in both the dispersion and energy of antiferromagnetic magnons has spurred the magnetism community to consider antiferromagnets for future spintronic/magnonic applications.  However, the excitation and control of antiferromagnetic magnons remains challenging, especially when compared to ferromagnetic counterparts.  A middle ground is found with synthetic antiferromagnet metamaterials, where acoustic and optical magnons exist at GHz frequencies.  In these materials, the magnon energy spectrum can be tuned by static symmetry-breaking external fields or dipolar interactions hybridizing  optical and acoustic magnon branches.  Here, we experimentally measure the magnon energy spectrum of synthetic antiferromagnetic \textit{tetralayers}, and discover avoided energy level crossings in the energy spectrum that are unexplained by the antiferromagnetic interlayer coupling.  We explain our experimental results using a phenomenological model incorporating both fieldlike and dampinglike torques generated by spin pumping in noncollinear magnetic multilayers separated by normal-metal spacers. We show that an asymmetry in the fieldlike torques acting on different magnetic layers can lift the spectral degeneracies of acoustic and optical magnon branches and yield symmetry-breaking induced magnon-magnon interactions. 
 Our work extends the phenomenology of spin pumping to noncollinear magnetization configurations and significantly expands ways of engineering magnon-magnon interactions within antiferromagnets and quantum hybrid magnonic materials.

\end{abstract}

\maketitle
Magnons are the low-energy quasiparticle correspondent of spin waves in magnetically ordered materials.  The coupling of magnons to other quasiparticles such as photons\cite{huebl2013high, zhang2014strongly, bai2015spin, harder2018level}, phonons\cite{weiler2011elastically, streib2019magnon, bozhko2020magnon}, and other magnons\cite{klingler2018spin, chen2018strong, qin2018exchange, xiong2020MIT,li2020coherent} has recently drawn great attention due to the emerging field of hybrid quantum systems.  The potential applications of such hybrid magnonic systems are vast\cite{li2020hybrid,awschalom2021quantum}.  One prospective application relies on on-chip magnon-photon interactions\cite{li2019strong, hou2019strong, baity2021strong,li2022coherent} to transduce information between qubits-to-photons-to-magnons\cite{tabuchi2015coherent,lachance2020entanglement}.  Antiferromagnetic materials, as a magnon platform, are particularly appealing because they possess both acoustic and optical magnons at small wavenumbers and can have frequencies at the GHz\cite{macneill2019gigahertz, waring2020zero}, sub-THz\cite{li2020spin,vaidya2020subterahertz}, and even THz level\cite{moriyama2019intrinsic,kampfrath2011coherent}.  In fact, this large frequency range has spurred the field of antiferromagnetic spintronics\cite{siddiqui2020metallic}, which aims to functionalize antiferromagnets as memories\cite{marti2014room, wadley2016electrical} and sources of THz radiation\cite{cheng2016terahertz,gomonay2018antiferromagnetic}.  Hybrid magnon materials, with tunable magnon-magnon interactions, offer new avenues for controlling antiferromagnetic magnon frequencies, making them novel candidate materials for spintronic applications\cite{li2020hybrid,awschalom2021quantum}.  The success of these future applications will depend on the discovery, characterization, and eventual control of magnon-magnon interactions within antiferromagnetic materials.


In this work, we use synthetic antiferromagnets (S-AFM) as a material platform to study hybridized magnons.  Conventional S-AFMs are comprised of ferromagnetic layers separated by normal metal spacer layers that promote an antiferromagnetic exchange-like coupling via the Ruderman–Kittel–Kasuya–Yosida (RKKY) interaction\cite{duine2018synthetic}, but can also be engineered to possess chiral exchange interactions\cite{yang2019chiral,han2019long} that can stabilize unconventional magnetization textures\cite{avci2021chiral} and skyrmions\cite{legrand2020room}.  In general, acoustic and optical magnons tend to exist at the single-GHz to the tens-of-GHz frequency scale within S-AFMs\cite{waring2020zero}.  Over this convenient range of frequencies, the magnon-magnon interaction between such modes has been previously tuned via the application of a symmetry-breaking external field\cite{sud2020tunable,li2021symmetry,dai2021strong, macneill2019gigahertz}.  Dipolar interactions were also shown to hybridize optical and acoustic magnons in a controlled manner when large enough magnon wavenumbers were excited\cite{shiota2020tunable}.  Recently, we predicted that the dynamic component of the RKKY exchange field could ``self-hybridize'' pairs of acoustic -or- optical magnons in layered antiferromagnets with four and six layers\cite{sklenar2021self}.  The former case, S-AFM tetralayers, are the main subject of investigation in this work.

When magnetization dynamics are driven coherently, interfacial spin pumping processes can also occur in S-AFMs. Early theoretical works have identified two types of spin pumping torques  in the Landau-Lifshitz-Gilbert (LLG) equations of motion governing coherently-driven  magnetizations dynamics, i.e., a dampinglike and a fieldlike torque associated with, respectively, the real and imaginary part of the spin-mixing conductance~\cite{tserkovnyak2002enhanced,brataas2012spin,heinrich2003dynamic,tserkovnyak2003dynamic}. The fieldlike torque is commonly neglected as it was suggested that the imaginary part of the spin-mixing conductance must be exceedingly smaller than its real part at  metallic interfaces. Here, we will  show that spin pumping fieldlike torques can instead play a dominant role in magnon band engineering.
Specifically, we will generalize the theory of interfacial spin pumping to noncollinear magnetization configurations and  we will demonstrate that, when the uppermost and the lowermost magnet$|$normal metal$|$magnet interfaces of tetralayer are not identical,  mode crossings between acoustic and optical branches of FMR spectra can be lifted by dynamical fieldlike torques.  The symmetry-breaking mechanism, to be elucidated below, that generates this new magnon-magnon interaction is illustrated in Figure 1.  In (a), an external field is applied to the tetralayer, and it is illustrated how the orientation of the effective magnetic fields that the surface (or interior) layers experience is mirror-symmetric about the external field direction.  As we will theoretically demonstrate, in the presence of any inversion asymmetry, dynamic interlayer fieldlike torques rotate the mirror-symmetry axes of the surface and interior effective fields away from one another [as shown in (b)].  In this scenario, we will experimentally show and theoretically describe how acoustic and optical magnons are subsequently able to interact.


Using DC magnetron sputtering (see Methods), we prepared both bilayer and tetralayer S-AFM films comprised of alternating permalloy/ruthenium (Py/Ru) layers that are grown between two platinum (Pt) layers that are 6 nm thick.    Our samples  exhibit antiferromagnetic interlayer coupling between 5 nm Py layers when the Ru layers are fixed at 1 nm.  The antiferromagnetic interaction is verified with magnetometry as seen in Figure 2 (a) and (b).  Antiferromagnetic samples have no magnetic hysteresis, and the magnetization increases with field linearly for the bilayer, and non-linearly for the tetralayer.  By fitting the normalized magnetization of our samples versus magnetic field (mathematical details are found in Supplemental material, Section 2.1)  we are able to extract an interlayer antiferromagnetic exchange field of $\mu_0H_E$ = 28.1 mT and $\mu_0H_E$ = 24.4 mT for the bilayer and tetralayer respectively.

To experimentally characterize the magnon energy spectra for both the bilayer and tetralayer we employ a broadband magnetic resonance technique where samples are flipped on top of a co-planar waveguide (CPW, see Methods).  In Figure 3, our most compelling results arise in an experimental geometry where samples are placed on the dielectric region between the signal and ground line of the CPW.  Here, the CPW generates an in-plane rf-field that is perpendicular to the signal line, as well as an an out-of-plane rf-field.  In Figure 3 (a) and (b), we show the experimentally excited magnon energy spectrum in the ``acoustic configuration'', where the external field is applied along the length of the signal line.  In this configuration, all dynamic components of the rf-field are perpendicular to the external field, and we only excite acoustic magnons.  For the bilayer, an unbroken linear magnon branch is observed, but for the tetralayer an avoided energy level crossing near 5.5 GHz breaks the main linear branch into two branches. 
 
When the external field is rotated to be perpendicular to the length of the signal line, it becomes parallel to the in-plane component of the rf-field.  We call this the ``optical configuration'' because optical magnons can now be excited by the in-plane component of the rf-field\cite{macneill2019gigahertz, sud2020tunable, shiota2020tunable}.  However, acoustic magnons are still excited due to the out-of-plane component of the rf-field.  The magnon spectrum in this configuration for the bilayer and tetralayer is shown in Figure 3 (c) and (d).  For the bilayer, we more strongly observe the low energy acoustic branch and do not couple to the optical magnon.  For the tetralayer we clearly observe \textit{three} magnon branches.  The lowest energy linear magnon branch is observed to dramatically flatten out as the external field increases.  A second, relatively flat, new magnon mode with a frequency of 4.9 GHz at zero field is observed.  Finally, a third magnon branch with a zero field frequency near 5.5 GHz is excited.  In addition to the surprising overall ``flatness'' of the two lower energy magnon branches, all three branches are experimentally observed to avoid one another.


To explain these  experimental measurements, we develop a general phenomenology for spin pumping in noncollinear magnetic layered structures incorporating both fieldlike and dampinglike spin pumping torques, consistent with Onsager reciprocity and symmetry principles. A detailed theory is provided in Supplemental material, Section 1. 
We  describe the magnetization dynamics of the tetralayer using four modified and coupled LLG equations:
\begin{align}
   \frac{d\mathbf{m}_{A(D)}}{dt} =&-\mu_{0}\gamma\mathbf{m}_{A(D)}\times \left[ H_{0} \hat{y} - H_{E} \mathbf{m}_{B(C)} - M_{s} \left( \mathbf{m}_{A(D)} \cdot \hat{z} \right) \hat{z} \right]\nonumber\\
+&\mathbf{m}_{A(D)}\times \left[\frac{d\mathbf{m}_{B(C)}}{dt}\times \left(\alpha_{f,AB(CD)}\mathbf{m}_{A(C)}+\beta_{f,AB(CD)}\mathbf{m}_{B(D)}\right)\right]\nonumber\\ +&\left(\alpha_{A(D)} + \alpha_{d,AB(CD)} \right)\mathbf{m}_{A(D)}\times \frac{d\mathbf{m}_{A(D)}}{dt}-\alpha_{d,AB(CD)}\mathbf{m}_{A(D)}\times\left(\mathbf{m}_{B(C)}\times\frac{d\mathbf{m}_{B(C)}}{dt}\right)\times\mathbf{m}_{A(D)}\,, \label{LLGAD} \\ \nonumber
    \frac{d\mathbf{m}_{B(C)}}{dt} =& -\mu_0\gamma \mathbf{m}_{B(C)} \times [H_0\hat{y} - H_E\left(\mathbf{m}_{A(B)} + \mathbf{m}_{C(D)}\right) -M_s(\mathbf{m}_{B(C)} \cdot \hat{z})\hat{z}]   \\ \nonumber+& \mathbf{m}_{B(C)}\times \left[\frac{d\mathbf{m}_{A(B)}}{dt}\times \left(\alpha_{f,AB(BC)}\mathbf{m}_{A(B)}+\beta_{f,AB(BC)}\mathbf{m}_{B(C)}\right)\right]  \\ \nonumber +& \mathbf{m}_{B(C)}\times \left[\frac{d\mathbf{m}_{C(D)}}{dt}\times \left(\alpha_{f,BC(CD)}\mathbf{m}_{B(C)}+\beta_{f,BC(CD)}\mathbf{m}_{C(D)}\right) \right] \\ \nonumber+&\left(\alpha_{B(C)} + \alpha_{d,AB(CD)}+\alpha_{d,BC}\right)\mathbf{m}_{B(C)}\times\frac{\mathrm{d}\mathbf{m}_{B(C)}}{\mathrm{d}t} -\alpha_{d,AB(BC)}\mathbf{m}_{B(C)}\times\left(\mathbf{m}_{A(B)}\times\frac{d\mathbf{m}_{A(B)}}{dt}\right)\times\mathbf{m}_{B(C)}\nonumber \\-&\alpha_{d,BC(CD)}\mathbf{m}_{B(C)}\times\left(\mathbf{m}_{C(D)}\times\frac{d\mathbf{m}_{C(D)}}{dt}\right)\times\mathbf{m}_{B(C)}\,. \label{LLGBC}
    \end{align}
where Eq.~(1) describes the dynamics of the magnetization on the top (bottom) surface layers $\mathbf{m}_{A(D)}$, while Eq. (2) represents the upper (lower) interior layers $\mathbf{m}_{B(C)}$.  $H_0$ and $H_E$ are the external and interlayer exchange fields, respectively.   The torque term proportional to the saturation magnetization, $M_s$, parameterizes the shape-dependent demagnetizing field of each thin film layer. The parameter $\alpha_{i}$ accounts for the local Gilbert damping of the $i$th layer, which we assume to be the same within each layer, i.e., $\alpha_{i}\equiv \alpha$, with $i=A,B,C,D$. Terms  $\propto \alpha_{d, ij}$ describe the dampinglike torques associated with spin pumping at the interface between the $i$th and $j$th layers and reduce to the well-known results of Ref.~\cite{heinrich2003dynamic} for a collinear magnetization configuration, i.e., $\mathbf{m}_{i} \parallel \mathbf{m}_{j}$.  
The terms $\alpha_{f,ij}$ and $\beta_{f,ij}$ parameterize the strength of the fieldlike torques at the interface between the $i$th and $j$th layers~\cite{Yaroslav}. In what follows, we treat each individual Py$|$Ru$|$Py bilayer as symmetric, which allows us to set $\alpha_{f,ij}=\beta_{f,ij}$ for a given $(i,j)$ pair of layers.
It is important to note that the dynamical fieldlike 
torques  contribute to the effective 
field $\mathbf{H}_{i,\text{eff}}$ acting 
on the magnetic order parameter 
$\mathbf{m}_{i}$, and thus influence the 
equilibrium orientation around which   the latter coherently precesses.  By solving self-consistently Eqs.~(\ref{LLGAD}) and~(\ref{LLGBC}), the effective field $\mathbf{H}_{i,f}$ acting on $\mathbf{m}_{i}$ due to spin pumping can be written, to first order, as
\begin{align}
\mathbf{H}_{A(D),f}=&\mathbf{m}_{B(C)} \times [H_0\hat{y} - H_E\left(\mathbf{m}_{A(B)}  + \mathbf{m}_{C(D)} \right) -M_s(\mathbf{m}_{B(C)} \cdot \hat{z})\hat{z}]
\times \alpha_{f,AB(CD)} \left(\mathbf{m}_{A(C)}+\mathbf{m}_{B(D)} \right)\,, \label{3}\\
\mathbf{H}_{B(C),f}=&\mathbf{m}_{A(D)}  \times [H_0\hat{y} - H_E\mathbf{m}_{B(C)}-M_s(\mathbf{m}_{A(D)} \cdot \hat{z})\hat{z}]\times \alpha_{f,AB(CD)}\left(\mathbf{m}_{A(C)}+\mathbf{m}_{B(D)} \right) \nonumber\\
&+\mathbf{m}_{C(B)} \times [H_0\hat{y} - H_E(\mathbf{m}_{B(C)}+\mathbf{m}_{D(A)})-M_s(\mathbf{m}_{C(B)} \cdot \hat{z})\hat{z}]\times \alpha_{f,BC}\left(\mathbf{m}_{B}+\mathbf{m}_{C}\right)\,. \label{4}
\end{align}
For finite spin pumping strength, i.e., $\alpha_{f,ij} \neq 0$, we find the (dynamical) equilibrium angles by minimizing the free energy while accounting for Eqs.~(\ref{3}) and~(\ref{4}), i.e., $\mathbf{H}_{i,\text{eff}} \rightarrow \mathbf{H}_{i,\text{eff}} + \mathbf{H}_{i,f}$.  To obtain the magnon energy spectrum we linearize the coupled LLG equations, seeking solutions of the form: $\mathbf{m}_{i} = (e^{i\omega t}\delta m_{i}^x, e^{i\omega t}\delta m_{i}^y,1) = (0,0,1)+\mathbf{\delta m}_{i}(t)$ in the local coordinates.  Here $\delta m_{i}^x$ and $\delta m_{i}^y$ are the small dynamic amplitudes that describe the elliptical precession of the magnetization around 
the direction of (dynamical) equilibrium.


To parse the different magnon-magnon interactions, we first discuss the magnon frequency-field dependence without spin pumping.  In this limit, Figures 4 (a) and (b) show that there are a pair of, respectively, acoustic and optical magnon branches.  In the absence of spin pumping, these ``pure'' optical and acoustic magnons can be differentiated in the theory (see Supplemental Section 2.2), and both the acoustic pairs and optical pairs tend to ``self-hybridize'' with one another due to the RKKY interaction $\propto H_{E}$. This self-hybridization generates two avoided energy level crossings, one for each pair of modes.  Previously, we simulated where the magnons spatially reside within the tetralayer\cite{sklenar2021self}.  Near the avoided energy level crossing, the low-energy acoustic magnon branch resides on the interior layers, and the low-energy optical magnon branch primarily resides on the surface layers.  In contrast, the high-energy acoustic magnon branch resides on the surface layers, and the high-energy optical magnon branch resides on the interior layers.  Thus, even if only considering the RKKY interaction, pairs of magnons sharing the same character (acoustic or optical) are found to interact with one another.  In Figure 4 (c), we plot the combined magnon spectrum, and we observe three level crossings, i.e., between acoustic I and optical I, optical I and acoustic II, acoustic II and optical II, and no additional interactions between any acoustic-optical magnon pair.

By comparing the experimental results from Figure 3 (b) with Figure 4 (a), we see agreement between experiment and theory in that the acoustic gap is centered near 5.5 GHz, with comparable gap sizes between theory and experiment.  Thus, when experimentally driving the tetralayer in the ``acoustic configuration'', the RKKY interaction is responsible for the avoided energy level crossing we observe.  In contrast, when comparing the theoretical spectrum shown in Figure 4 (c) with the experimentally measured spectrum in the ``optical configuration'' [Figure 3 (d)], it is clear that the model sans spin pumping is unable to reproduce (1) the flattening of the lowest magnon branch and (2) the mutual avoidance among all three magnon branches.  In order to explain these experimental observations we use the extended model, where we find that acoustic and optical modes interact via the dynamic fieldlike and dampinglike torques exerted on one layer by the neighboring layers via spin pumping.  The dampinglike torque amounts to a non-Hermitian contribution to the effective Hamiltonian derived upon linearization of Eqs.~(\ref{LLGAD}) and (\ref{LLGBC})~\cite{NonHermitianreview}. The latter manifests as a change in the magnon linewidth and as magnon level attraction~\cite{heinrich2003dynamic,yang2016large, sorokin2020magnetization, troncoso2021cross}, which can lead to energy coalescence and, thus, exceptional points~\cite{NonHermitianreview}. The fieldlike torques amount instead to additional fields  acting on the magnetization dynamics, i.e., see Eqs.~(\ref{3}) and~(\ref{4}), leading to level repulsion between acoustic and optical modes.

In order to reproduce the flattening of the low energy magnon branch as well as the mutual avoidance amongst all three experimentally observed magnon branches, we set $\alpha_{f,AB}=\alpha_{f,BC}=0.1$ and $\alpha_{f,CD}=0.15$.   We set the dampinglike terms as $\alpha_{d,AB}=\alpha_{d,BC}=0.01$, and $\alpha_{d,CD}=0.015$, noting that a smaller magnitude in the damping like torques is needed to best reproduce experiment.  The local Gilbert damping is set to $\alpha=0.01$ for our calculations.  In Figure 4 (d) the resulting magnon branches are plotted.  As indicated by dashed boxes, the original level crossings have turned into avoided level crossings, thereby creating mutual avoidance between the three lowest magnon branches.  Furthermore, the two lowest energy magnon branches flatten out as the field increases, which.  Thus, the measured experimental spectrum in the optical configuration shown in Figure 3 (d) is most consistent with calculations incorporating interlayer fieldlike torques arising from interlayer spin pumping.

The torque parameters best describing our experimental results imply that an asymmetry exists between 
the top and bottom interfaces, i.e., $\alpha_{f,AB}\neq \alpha_{f,CD}$.  This asymmetry is 
responsible for lifting the degeneracies of the magnon spectrum and opening energy gaps between 
optical and acoustic magnons; an extended discussion can be found in Supplemental material, Section 
1.  The consequence of this asymmetry, illustrated in Figure 1, is that the symmetry axes describing 
the effective fields the surface and interior layers experience rotates away from the external field 
direction in opposite directions, i.e., $\mathbf{H}_{A,\text{eff}}\neq\mathcal{C}_{2y} 
\mathbf{H}_{D,\text{eff}}$ and $\mathbf{H}_{B,\text{eff}} \neq \mathcal{C}_{2y} 
\mathbf{H}_{C,\text{eff}}$.  Evidence for inversion asymmetry in the structure of the sample itself is found by a series of atomic force microscopy measurements shown in the Supplemental material, Section 5.  Atomic force microscopy characterization of the surface roughness before and after metal 
deposition shows that the overall surface roughness decreases from 203 pm to 157 pm (averaged over 
three samples).  The  reduction in the surface roughness as the film thickness increases suggests 
that the ruthenium interlayers the interface between the ruthenium and permalloy at the bottom of the 
tetralayer is not equivalent to the interface at the top of the tetralayer due to the substrate.



An alternative way we characterized the magnon energy spectrum of the tetralayer was by using micro-focused Brillouin light spectroscopy (BLS) to measure the incoherent thermal magnon spectrum.  In contrast to the coherent spectrum, where the rf-field excites magnons in the long wavelength limit, the BLS spectrum is integrated over a range of wavenumbers ($0- 17.8$ $ rad/\mu m$) and wavevector orientations relative to the external field.  In Figure 5, the thermal spectrum shows a low energy acoustic magnon branch as well as an optical magnon branch that has a zero-field frequency just below 5 GHz. However, the low energy acoustic branch is not observed to flatten as the field is increased.  Additionally, no avoided energy level crossings are conclusively resolved between the two observed magnon branches.  In the region where the avoided energy level crossing is observed in the coherent exciation scheme, the thermal magnon spectrum has a strong thermal magnon signal.  This result suggests that there is a non-trivial wavevector dependence on the magnon-magnon interactions within the tetralayer.  To date, wavenumber dependent measurements of magnon-magnon interactions in S-AFMs are limited.  Only Shiota \textit{et al.} have shown how magnon-magnon interactions, mediated by the dipolar coupling of magnons, can be enhanced by the wavevector orientation in S-AFM bilayers\cite{shiota2020tunable}.  Our results suggests that the dependence of magnon-magnon interactions mediated by symmetry breaking fields, dipolar interactions, and spin pumping is a future direction worth of exploration an even finer control of magnonic interactions within magnetic metamaterials.

In this work, the oft-neglected dynamic fieldlike torques generated by spin pumping were able to hybridize acoustic and optical magnons in S-AFM tetralayers. Looking ahead, the flexibility of the S-AFM material platform should allow for alternative multilayer structures where spin pumping can be either suppressed or enhanced, which leads to a new manner in which magnons can be manipulated and tuned in synthetic magnets.  The phenomenology we have developed is not limited to just tetralayers, and can be applied to any other noncollinear magnetic multilayer structure.  By using this phenomenology to describe other multilayers with specifically defined interfaces, we suggest that synthetic antiferromagnets are a playground not just for their flexibility in allowing experimentalists to control magnon-magnon interactions, but also for non-Hermitian phenomena.  Future experiments that are able to harness dynamic antidamping torques associated with spin pumping may provide the key in unlocking the experimental ability to engineer exceptional points directly into the energy spectrum of synthetic antiferromagnetic materials.

\section{Acknowledgements}
We thank Yaroslav Tserkovnyak, Axel Hoffmann and Eric Montoya for insightful and valuable discussions during the preparation of this manuscript. 

Work at Oakland University was supported by  U.S.  National  Science  Foundation  under  award  No. ECCS-1941426.  P.B.M and J.T.H were supported from  NSF CAREER grant DMR-1847847.  Research was supported by NSF through the University of Delaware Materials Research Science and Engineering Center, DMR-2011824. B.F. was supported by the NSF under Grant No. NSF DMR-2144086.   The authors acknowledge the use of facilities and instrumentation supported by NSF through the University of Delaware Materials Research Science and Engineering Center, DMR-2011824.

\section{Methods}
All the samples that were used for magnon measurements were deposited by DC magnetron sputtering with argon
plasma onto double sided polished sapphire substrates at room temperature. Sputtering is performed in a high vacuum system with a base pressure near $3 \times 10^{ - 9}$ Torr. During deposition, a 3 mTorr atmosphere of argon gas is introduced into the chamber and is ignited into a plasma.  We synthesized two series of tetralayer samples, Pt/Py/Ru/Py/Ru/Py/Ru/Py/Pt and Ru/Py/Ru/Py/Ru/Py/Ru/Py/Pt.  The top and bottom layers of the overall structure consist of either Pt or Ru.  We found that by encapsulating the Py/Ru layers within such a structure, we suppressed any residual ferromagnetic moment from forming due to uncompensated magnetization.  The top and bottom Pt layers were fixed to be red 6 nm, the Py layers were fixed to be 5 nm, and we varied the thickness of the Ru layers from 0.25 nm to 3 nm with equal interval of 0.25 nm.  The DC power used in the growth of Py and Ru layers is 50 W, and the DC power is 30 W for Pt.

\section{Characterization}
Tetralayer samples are placed on a broadband coplanar waveguide (CPW)  upside down, in a so-called flip-chip configuration.  One side of the CPW is connected to a microwave signal generator (1-20 GHz range), and the other side is connected to a microwave diode.  The output of the diode is connected to the input of a lock-in amplifier.  The CPW sample assembly is then placed half-way between Helmholtz coils and two pole pieces of electromagnet.  The electromagnet is used to sweep the external field, and the Helmholtz coils are connected to an audio amplifier that is driven at the reference frequency of the lock-in amplifier (103.73 Hz).  We use the Helmholtz coils to modulate the swept external field with an amplitude of about 8 Oe, so that we can measure the transmitted output signal of the diode with the lock-in amplifier.  Both the in-phase component (X-value) and out-of-phase component (Y-value) of the output of microwave diode are recorded corresponding to the relative phase difference between the measurement signal and reference signal.

Microfocused Brillouin light scattering spectroscopy measurements are performed in a backscattering geometry using a continuous wavelength single-mode 532 nm wavelength laser. A microscope objective with a magnification of 100x, a high numerical aperture of 0.75, and a working distance of 4 mm is used to focus the laser beam on the sample surface. An additional light source is used to illuminate the sample enabling probing position control during the measurements. An autofocus routine is implemented by monitoring the reflected laser intensity from the sample by a photodiode. The inelastically scattered light is analyzed using a high-contrast multi-pass tandem Fabry Pérot interferometer with a contrast of at least 1015. The magnon signal is extracted from the Stokes peak unless stated otherwise. All BLS measurements are conducted at room temperature without any microwave excitation; therefore, the detected signal corresponds to the thermal magnon spectrum in the tetralayers.


\newpage

\begin{figure*}
    \centering
    \includegraphics[scale=0.23]{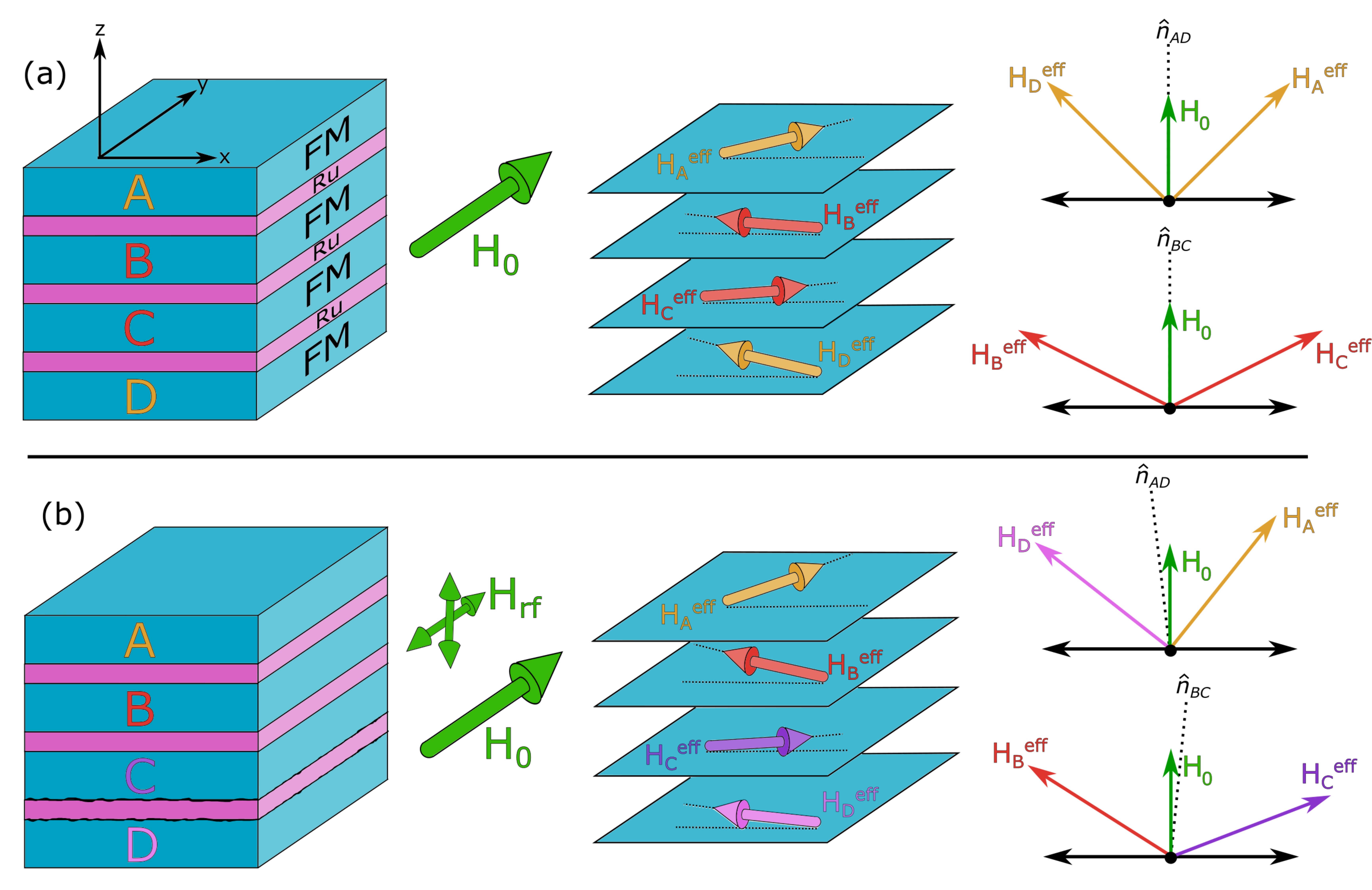}
    \caption{\textbf{Spin pumping-driven symmetry breaking in a synthetic antiferromagnet.}  In (a), a S-AFM tetralayer is illustrated.  Tetralayers consist of four ferromagnetic (FM) layers separated by three Ru interlayers.  Surface layers are labeled \textit{A} $\&$ \textit{D}, while interior layers are labeled \textit{B} $\&$ \textit{C}.  (a) continues to illustrate the conventional magnetostatic picture where, if  an external field is applied along the $y$-direction, the effective magnetic field directions that the surface layers experience ($H_A^{eff}$ $\&$ $H_D^{eff}$) are mirror symmetric about the $y$-axis.  Similarly, the effective fields experience by the interior layers ($H_B^{eff}$ $\&$ $H_C^{eff}$) are also mutually mirror symmetric about the external field direction.  In (b), we consider a more realistic situation where there is some asymmetry between the metallic interface separating \textit{A} $\&$ \textit{B}, compared to the interface separating \textit{C} $\&$ \textit{D}.  In this case, when both an external field and driving rf-field are present, interlayer spin pumping can break the symmetry of the system.  This broken symmetry is further illustrated in the rightmost panel of (b) where the mirror symmetry axes of the surface layers and interior layers ($\hat{n}_{AD}$ $\&$ $\hat{n}_{BC}$) rotate in opposite directions away from the external field direction.   }
    \label{fig:my_label}
\end{figure*}

\begin{figure}
\includegraphics[scale = .35]{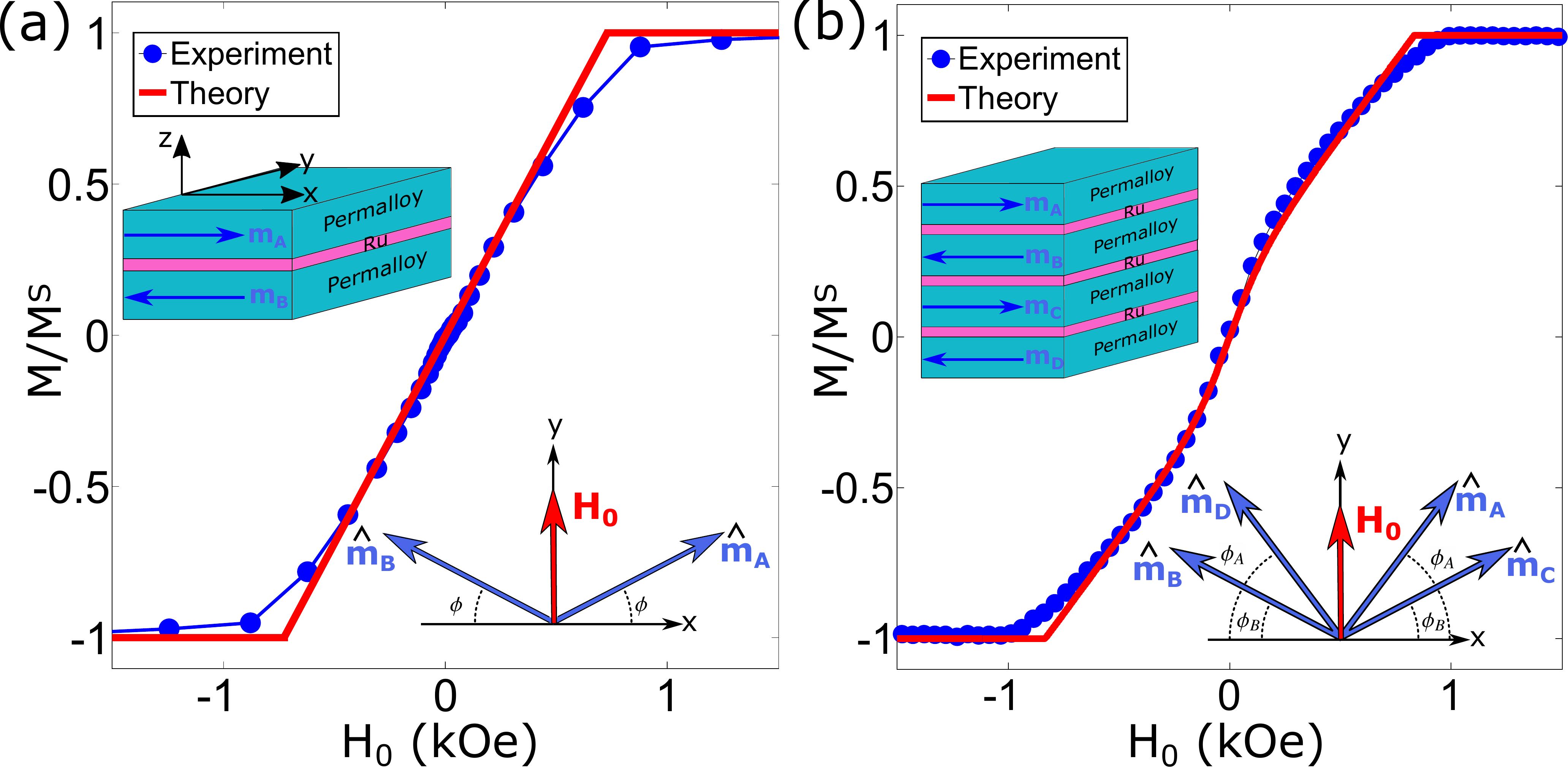}
\caption{\textbf{Static magnetic properties of synthetic antiferromagnetic bilayers and tetralayers.}  (a) The normalized magnetization of the S-AFM bilayer is plotted as a function of the external field applied in the plane of the sample.  The inset images illustrate the bilayer structure as well as the in-plane magnetization configuration which can be described by two unit vectors ($\hat{m}_A$ and $\hat{m}_B$) as well as a single angle ($\phi$) which describes the rotation of the magnetization of each layer towards the external field.  (b) The normalized magnetization of the S-AFM tetralayer is plotted as a function of the external field applied in the plane of the sample.  The inset images illustrate the tetralayer structure as well as the in-plane magnetization configuration which can be described by four unit vectors ($\hat{m}_{A(B)(C)(D)}$) as well as two angles ($\phi_A$ and $\phi_B$) that describes the rotation of the magnetization of the surface layers and interior layers towards the external field.  The field dependence of the equilibrium angles can be theoretically modeled (see Supplemental Section 2) to describe the magnetization curves and extract the interlayer exchange field $H_E$. }
\label{fig:MPMS}
\end{figure}

\begin{figure}
\includegraphics[scale = .2]{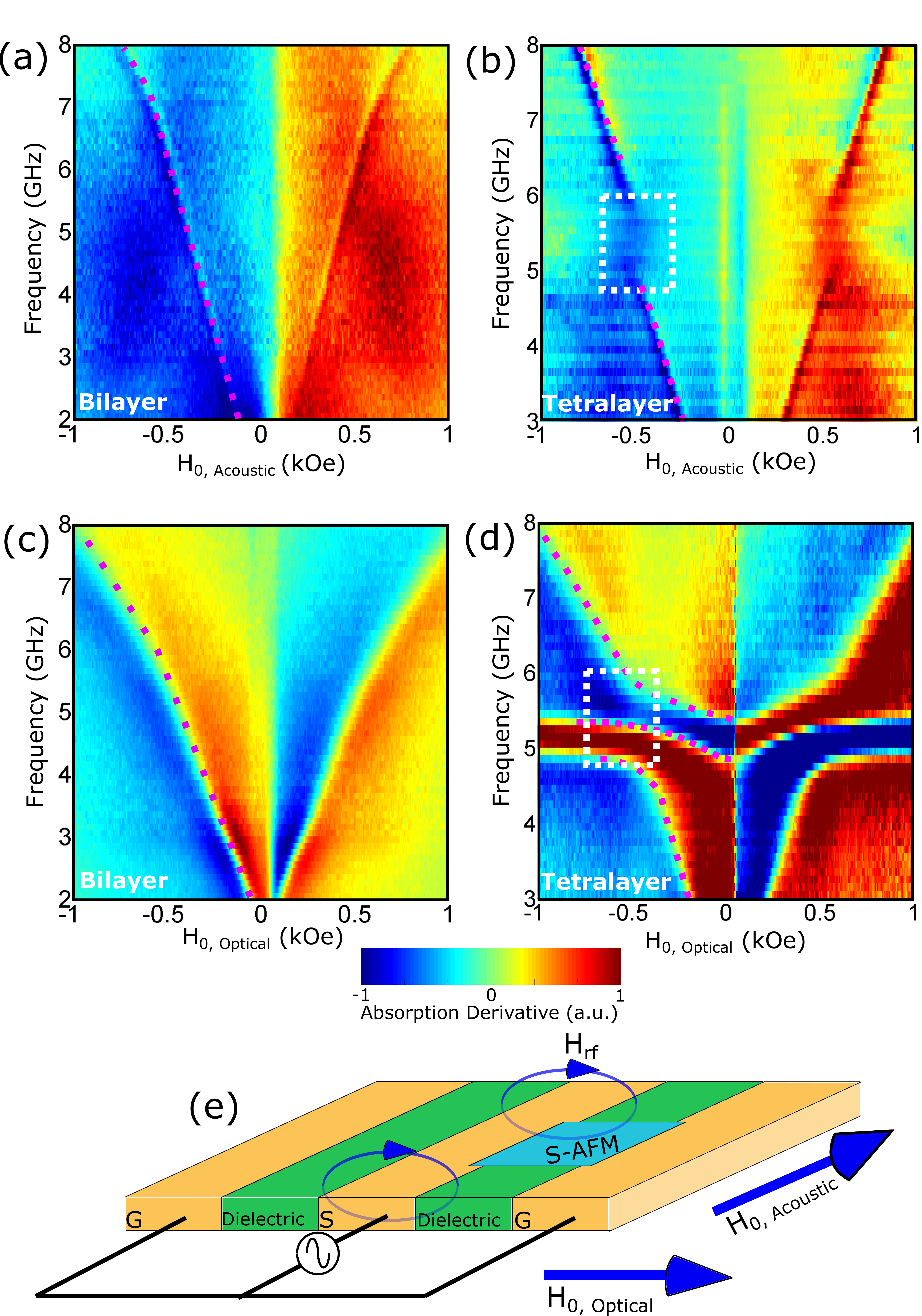}
\caption{\textbf{Experimentally measured magnon spectra.}  In (a) and (b) the field-frequency dependence on the derivative of the magnetic resonance absorption signal is plotted using a color scale for the bilayer and tetralayer.  Here, the external field is applied to be perpendicular to all the rf-fields driving the magnetization dynamics from the co-planar waveguide.  This is the ``acoustic'' configuration illustrated in (e).  Color contrast, denoted by the dashed lines, indicate where the magnons are excited.  A single unbroken magnon branch, symmetric upon reversal of the external field, is seen for the bilayer.  A clear avoided energy level crossing, breaking the spectrum into two branches is seen for the tetralayer. In (c) and (d) the same field-frequency dependence is plotted except that the external field is rotated such that the in-plane component of the rf-field is now parallel to the external field.  This is the ``optical'' configuration illustrated in (e).  Color contrast, denoted by the dashed lines, indicate where the magnons are excited.  A single unbroken magnon branch, symmetric upon reversal of the external field, is still seen for the bilayer.  In the tetrlayer, three magnon branches, helpfully denoted by dashed lines are excited; all three branches mututally avoid one another.  The dashed white box denotes the region where all three branches clearly avoid each other.}
\label{fig:FMR}
\end{figure}

\begin{figure*}
\includegraphics[width=\linewidth]{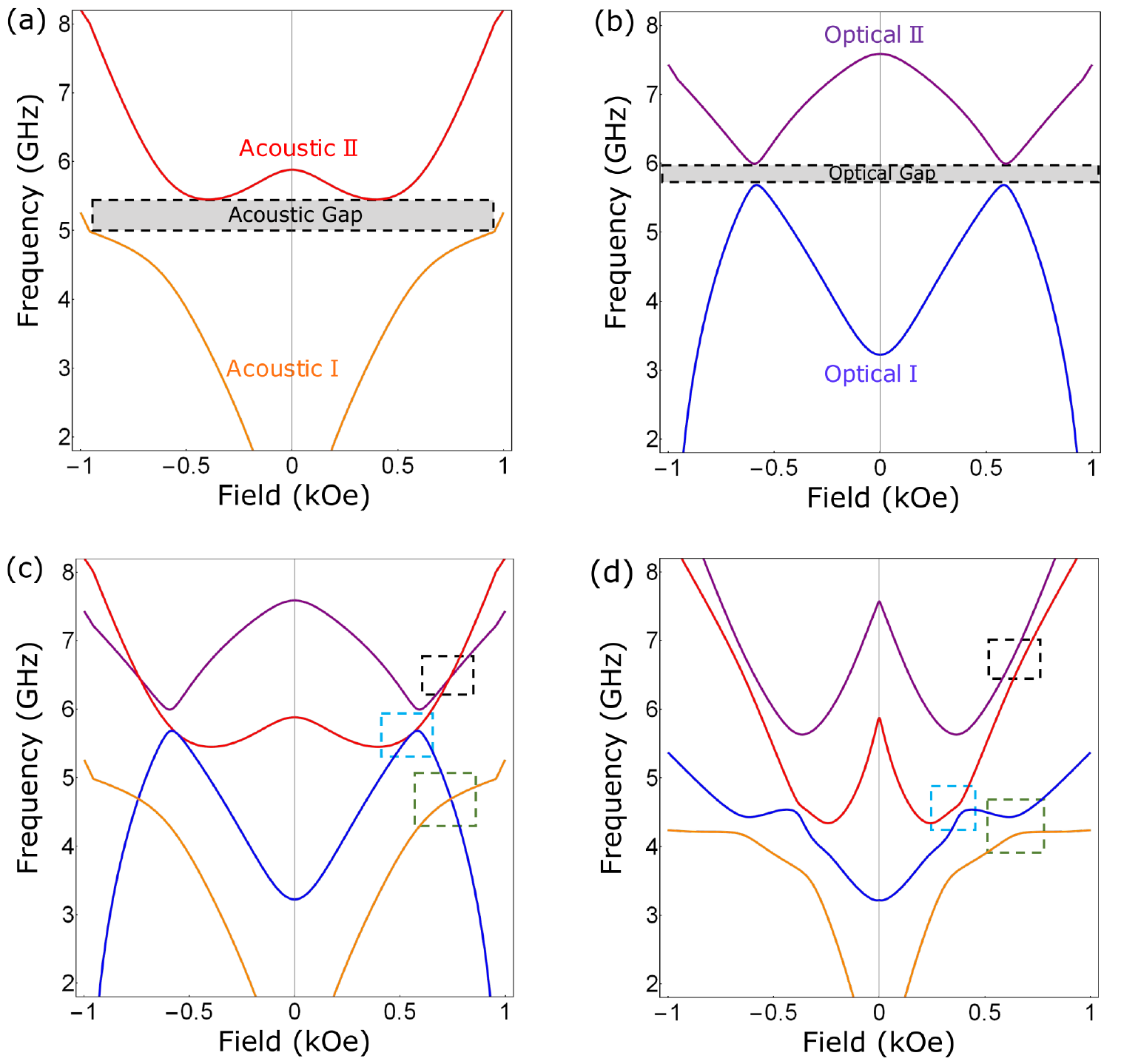}
\caption{{\textbf{Theoretically predicted acoustic and optical magnon spectra, including magnon-magnon interactions.} (a), (b), and (c) represent the acoustic, optical, and combined magnon spectrum in the absence of spin pumping for the tetralayer.  Due to the dynamic exchange field, the two acoustic magnons (a), and the two optical magnons (b), ``self-hybridize'' with one another.  This creates the acoustic and optical gap that is indicated in (a) and (b), respectively. In (c), we observe that no interactions exist between optical and acoustic magnons, as seen in the field-frequency relationship regions highlighted by the dashed boxes. In (d) we calculate the magnon spectrum of the tetralayer for $\alpha_{f,AB}=\alpha_{f,BC}=0.1$, $\alpha_{f,CD}=0.15$, $\alpha_{d,AB}=\alpha_{d,BC}=0.01$, $\alpha_{d,CD}=0.015$ and $\alpha=0.01$. Due to the asymmetry between the  uppermost and the lowermost interface, i.e., $\alpha_{f,AB}\neq \alpha_{f,CD}$, the three energy-level crossings between acoustic and optical magnon branches are lifted, and new branches emerge as the result of the interface-driven magnon-magnon hybridization.
}}
\label{fig:THEORY}
\end{figure*} 

\begin{figure}
\includegraphics[scale = .35]{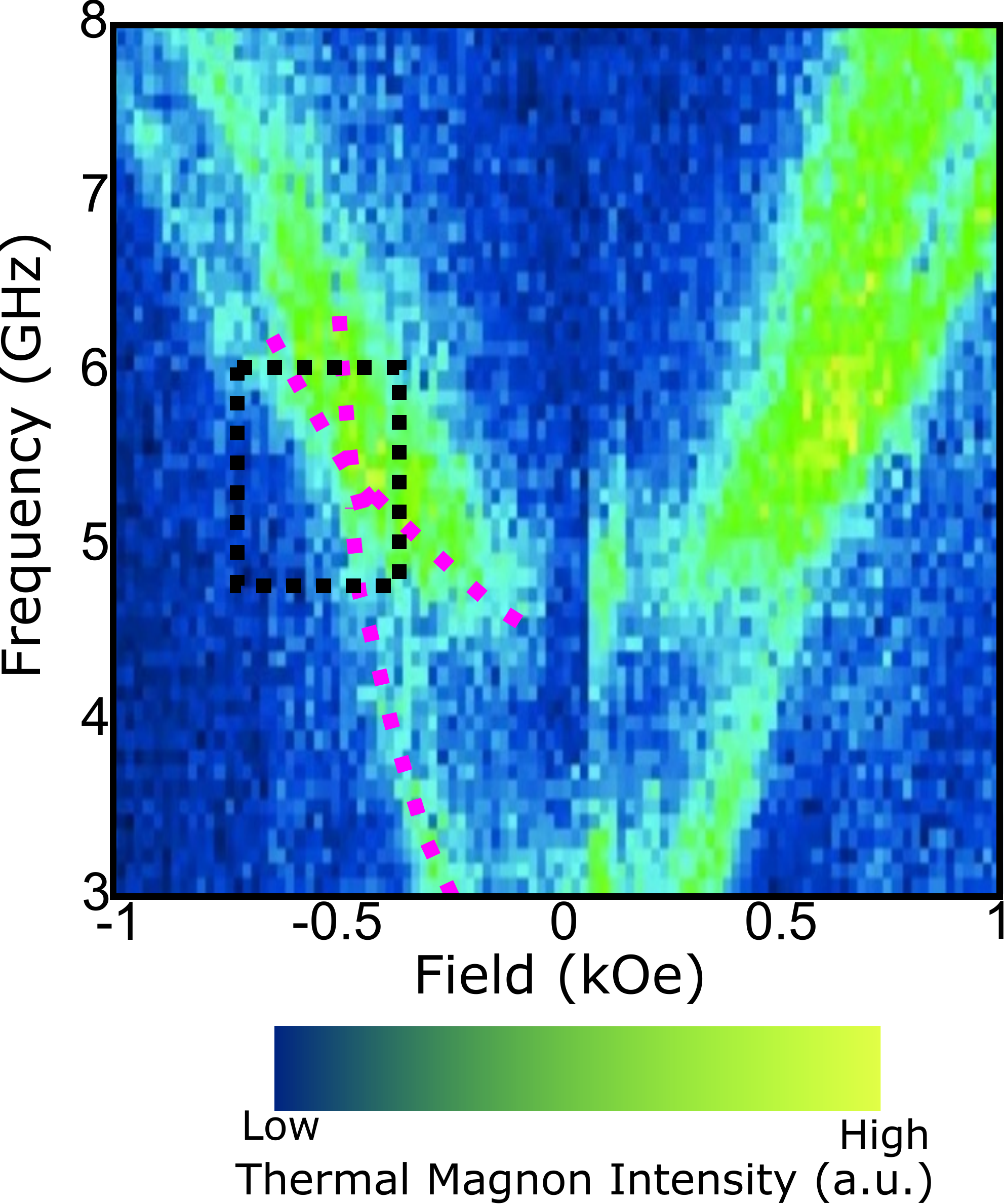}
\caption{\textbf{Incoherent thermal magnon spectrum of the tetralayer.} The field-frequency dependence of incoherent thermal magnons are detected at room temperature using microfocused Brillouin light scattering spectroscopy.  The color scale shows (in arbitrary units) where the magnon intensity is high (bright green) and low (dark blue).  Only two magnon branches are observed, and pink dashed lines serve as guides to the eye where the two meet.  Note that the lowest energy branch does not appear to flatten as the external field is increases, nor does there appear to be any clear avoided energy level crossing. }
\label{fig:BLS}
\end{figure}

\end{document}


\maketitle

\section{Spin pumping in noncollinear magnetic bilayers}

In this section, we derive phenomenological expressions for the fieldlike and dampinglike torques resulting from spin pumping between two noncollinear ferromagnetic films separated by a normal-metal spacer. We start by introducing $\textbf{m}_{A}$ and $\textbf{m}_{B}$ as the magnetization orientations of the ferromagnetic layer $A$ and $B$, respectively. In terms of these variables, the instantaneous state of the bilayer can be described by a free-energy functional $\mathcal{F}[\mathbf{m}_{A},\mathbf{m}_{B}$], and   the  (Landau-Lifshitz) transverse field acting on the magnetic order parameter $\mathbf{m}_{A(B)}$ is defined as $\mathbf{H}_{A(B)}\equiv \delta_{\mathbf{m}_{A(B)}} \mathcal{F}[\mathbf{m}_{A},\mathbf{m}_{B}]$. Within linear response, the relations between the rates $\dot{\mathbf{m}}_{A}$ and $\dot{\mathbf{m}}_{B}$ can be written as
\begin{align}
\begin{pmatrix} \dot{\mathbf{m}}_{A} \\ \dot{\mathbf{m}}_{B} \end{pmatrix}=\begin{pmatrix} L^{\mathbf{m}_{A} \mathbf{m}_{A}}   & L^{\mathbf{m}_{A} \mathbf{m}_{B}} \\  L^{\mathbf{m}_{B} \mathbf{m}_{A}} & L^{\mathbf{m}_{B} \mathbf{m}_{B}} \end{pmatrix} \begin{pmatrix} \mathbf{H}_{A} \\ \mathbf{H}_{B} \end{pmatrix}\,,
\end{align}
where $\mathbf{L}$ is a 4x4 linear response matrix, whose form is restricted by Onsager reciprocity~\cite{onsager}, i.e., 
\begin{align}
    \left[ L^{\mathbf{m}_{A(B)}\mathbf{m}_{A(B)}}\left( \mathbf{m}_{A(B)} \right)\right]_{ij}&=\left[L^{\mathbf{m}_{A(B)}\mathbf{m}_{A(B)}}\left( -\mathbf{m}_{A(B)} \right)\right]_{ji}\,, \\ 
    \left[ L^{\mathbf{m}_{A}\mathbf{m}_{B}}\left( \mathbf{m}_{A}, \mathbf{m}_{B} \right)\right]_{ij}&=\left[L^{\mathbf{m}_{B}\mathbf{m}_{A}}\left( - \mathbf{m}_{A}, -\mathbf{m}_{B} \right)\right]_{ji}\,. \label{Onsager}
\end{align}
The 2x2 block matrix $L^{\mathbf{m}_{A(B)}\mathbf{m}_{A(B)}}$ describes the decoupled Landau-Lifshitz-Gilbert dynamics of the order parameter $\mathbf{m}_{A(B)}$, whose dissipative component is accounted for via the Gilbert damping parameter $\alpha_{A(B)}$. The 2x2 block matrix $L^{\mathbf{m}_{A(B)}\mathbf{m}_{B(A)}}$ accounts for the spin pumping processes between the coherently-driven magnetizations dynamics.
Let us start by writing the coupled equations of motion in a general form as
\begin{align}
\label{38}
\frac{d\mathbf{m}_{A}}{dt} =& -\mu_{0}\gamma \mathbf{m}_{A} \times \mathbf{H}_{A} + \alpha_{A} \mathbf{m}_{A} \times \frac{d\mathbf{m}_{A}}{dt} + \alpha_{d,A} \mathbf{m}_{A} \times \left[ \mathbf{m}_{A} \times \frac{d\mathbf{m}_{A}}{dt} - \mathbf{m}_{B} \times \frac{d\mathbf{m}_{B}}{dt} \right] \times \mathbf{m}_{A} \\ \nonumber +& \mathbf{m}_{A} \times \left[ \frac{d\mathbf{m}_{B}}{dt} \times \left( \alpha_{f,A} \mathbf{m}_{A} + \beta_{f,A} \mathbf{m}_{B} \right) \right] \,, 
\end{align}
\begin{align}
\label{39}
\frac{d\mathbf{m}_{B}}{dt} =& -\mu_{0}\gamma \mathbf{m}_{B} \times \mathbf{H}_{B} + \alpha_{B} \mathbf{m}_{B} \times \frac{d\mathbf{m}_{B}}{dt} + \alpha_{d,B} \mathbf{m}_{B} \times \left[ \mathbf{m}_{B} \times \frac{d\mathbf{m}_{B}}{dt} - \mathbf{m}_{A} \times \frac{d\mathbf{m}_{A}}{dt} \right] \times \mathbf{m}_{B} \\ \nonumber +& \mathbf{m}_{B} \times \left[ \frac{d\mathbf{m}_{A}}{dt} \times \left( \alpha_{f,B} \mathbf{m}_{B} + \beta_{f,B} \mathbf{m}_{A} \right) \right]\,.
\end{align}
Invoking Onsager reciprocity~(\ref{Onsager}), we find $\alpha_{d,A}=\alpha_{d,B}$, $\alpha_{f,A}=\alpha_{f,B}$ and $\beta_{f,A}=\beta_{f,B}$. Here, $\alpha_{d,AB} \equiv \alpha_{d,A(B)} \ll 1$  is a phenomenological  coefficient parameterize the strength of the dampinglike spin-pumping torques. The terms $\propto \alpha_{f,AB} \equiv \alpha_{f,A(B)}$ and  $\propto \beta_{f,AB} \equiv \beta_{f,A(B)}$, with $\alpha_{f,AB},\beta_{f,AB} \ll 1$,  describe the fieldlike torques associated with spin pumping processes. 
Equations~(\ref{38}) and~(\ref{39}) can be rewritten as
\begin{align}
\label{LLGfinal1}
\frac{d\mathbf{m}_{A}}{dt} =& -\mu_{0}\gamma \mathbf{m}_{A} \times \mathbf{H}_{A} + \left(\alpha_{A} + \alpha_{d,AB} \right) \mathbf{m}_{A} \times \frac{d\mathbf{m}_{A}}{dt} - \alpha_{d,AB} \mathbf{m}_{A} \times \left(   \mathbf{m}_{B} \times \frac{d\mathbf{m}_{B}}{dt} \right) \times \mathbf{m}_{A} \\ \nonumber +& \mathbf{m}_{A} \times \left[ \frac{d\mathbf{m}_{B}}{dt} \times \left( \alpha_{f,AB} \mathbf{m}_{A} + \beta_{f,AB} \mathbf{m}_{B} \right) \right]\,, 
\end{align}
\begin{align}
\label{LLGfinal2}
\frac{d\mathbf{m}_{B}}{dt} =& -\mu_{0}\gamma \mathbf{m}_{B} \times \mathbf{H}_{B} + \left(\alpha_{B} + \alpha_{d,AB} \right) \mathbf{m}_{B} \times \frac{d\mathbf{m}_{B}}{dt}  - \alpha_{d,AB} \mathbf{m}_{B} \times \left(  \mathbf{m}_{A} \times \frac{d\mathbf{m}_{A}}{dt} \right) \times \mathbf{m}_{B} \\ \nonumber +& \mathbf{m}_{B} \times \left[ \frac{d\mathbf{m}_{A}}{dt} \times \left( \alpha_{f,AB} \mathbf{m}_{B} + \beta_{f,AB} \mathbf{m}_{A} \right) \right]\,.
\end{align}

Introducing the parity operation as the interchange $\mathbf{m}_{A} \leftrightarrow \mathbf{m}_{B}$, which implies $\mathbf{H}_{A} \leftrightarrow \mathbf{H}_{B}$,  one can easily see from Eqs.~(\ref{LLGfinal1}) and~(\ref{LLGfinal2})  that, for a symmetric bilayer, one has $\alpha_{f,AB}=\beta_{f,AB}$ and $\alpha_{A}=\alpha_{B}$. Figs, ~S1 (c) and (d) show that the fieldlike torques can lift the degeneracy between the acoustic and the optical mode  upon parity symmetry breaking, i.e., $\alpha_{f,AB} \neq \beta_{f,AB}$.  Similarly, we find that the energy coalescence in parameter space increases only when $\alpha_{A} \neq \alpha_{B}$, as shown by Figs.~S1 (a) and (b).

\section{Tetralayer model}

\subsection{Linearization of the equations of motion}

We model the magnetization dynamics of a synthetic antiferromagnetic tetralayer by assuming that the magnetic order parameter $\mathbf{m}_{i}$ of the layer $i$, with $i=A,B,C,D$, is statically and dynamically coupled to its nearest neighbors. In the short-range coupling regime, it is straightforward to generalize Eqs.~(\ref{LLGfinal1}) and~(\ref{LLGfinal2}) to Eqs.~(1) and~(2) of the main text. To solve for the magnon eigenfrequencies, we have to linearize the coupled LLG equations~(1) and~(2). For each layer, we can orient a spin-space Cartesian coordinate system such that the $\hat{z}$ axis locally lies along the classical orientation of the magnetic order parameter $\mathbf{m}'_{i}$. The latter can be related to order parameter $\mathbf{m}_{i}$ in the global frame of reference via the transformation
\begin{align}
\mathbf{m}_{i}=\mathcal{R}_{iG} \mathbf{m}'_{i}\,, \; \; \; \; \text{with}\; \;\mathcal{R}_{iG}\equiv \mathcal{R}_{z}(\phi_{i}) \mathcal{R}_{y}(\varphi_{i})\,,
\end{align}
where the matrix $\mathcal{R}_{z(y)}(\theta)$ describes a right-handed rotation by an angle $\theta$ about the $\hat{z}$ $    (\hat{y})$ axis, and  $\varphi_i$ ($\phi_i$) is the polar (azimuthal) angle of the equilibrium orientation of the order parameter $\mathbf{m}_{i}$. The magnetic order parameter $\mathbf{m}'_{i}$ can be written in terms of its static and dynamical components as
\begin{align*}
    \mathbf{m}'_{i} = \hat m_{i,eq}+\mathbf{\delta m}_i,
    \label{73}
\end{align*}
with $\hat m_{i,eq}=(0,0,1)$ and $\mathbf{\delta m}_i=e^{i\omega t}(\delta m_i^x,\delta m_i^y, 0)$, with $|\delta m^{x(y)}_{i}| \ll 1$. 
One can then rewrite Eqs.~(1) and (2) in the local frame as
\begin{align}
i\omega\delta\mathbf{m}_{A(D)}=&-\mu_{0}\gamma\mathbf{m}_{A(D)}\times [ H_{0} \mathcal{R}_{GA(D)} \hat{y} - H_{E} \mathcal{R}_{BA(CD)} \mathbf{m}_{B(C)} - M_{s} \left( \mathbf{m}_{A(D)} \cdot \mathcal{R}_{GA(D)}\hat{z} \right) \mathcal{R}_{GA(D)}\hat{z} ]\nonumber
 \\ &-\mu_{0} \gamma \mathbf{m}_{A(D)}\times \mathcal{R}_{GA(D)} \mathbf{H}_{A(D),f} \nonumber +  i\omega\left(\alpha_{A(D)} + \alpha_{d,AB(CD)} \right) \mathbf{m}_{A(D)}\times \mathbf{\delta m}_{A(D)}  \\ & - i\omega\alpha_{d,AB(CD)}\mathbf{m}_{A(D)}\times\left(\mathcal{R}_{BA(CD)}\mathbf{m}_{B(C)}\times \mathcal{R}_{BA(CD)}\mathbf{\delta m}_{B(C)} \right)\times\mathbf{m}_{A(D)} \\
i\omega\delta\mathbf{m}_{B(C)}=&-\mu_{0}\gamma\mathbf{m}_{B(C)}\times [ H_{0} \mathcal{R}_{GB(C)} \hat{y} - H_{E}\left(  \mathcal{R}_{AB(DC)} \mathbf{m}_{A(D)}+\mathcal{R}_{CB(BC)} \mathbf{m}_{C(B)} \right) \nonumber
\\& -M_{s} \left( \mathbf{m}_{B(C)} \cdot \mathcal{R}_{GB(C)}\hat{z} \right) \mathcal{R}_{GB(C)}\hat{z}]
-\mu_{0} \gamma \mathbf{m}_{B(C)}\times \mathcal{R}_{GB(C)} \mathbf{H}_{B(C),f}\nonumber\\
&+ i\omega \left(\alpha_{B(C)} + \alpha_{d,AB(CD)}+\alpha_{d,BC}\right) \mathbf{m}_{B(C)}\times \mathbf{\delta m}_{B(C)}  \nonumber\\ 
& -i\omega\alpha_{d,AB(BC)}\mathbf{m}_{B(C)}\times\left(\mathcal{R}_{AB(BC)}\mathbf{m}_{A(B)}\times \mathcal{R}_{AB(BC)}\mathbf{\delta m}_{A(B)} \right)\times\mathbf{m}_{B(C)} \nonumber\\ 
&  -i\omega\alpha_{d,BC(CD)}\mathbf{m}_{B(C)}\times\left(\mathcal{R}_{CB(DC)}\mathbf{m}_{C(D)}\times\mathcal{R}_{CB(DC)}\delta\mathbf{m}_{C(D)}\right)\times\mathbf{m}_{B(C)},, 
\end{align}
where we have introduced $\mathcal{R}_{ij}=\mathcal{R}_{jG}^{-1}\mathcal{R}_{iG}$ and the fields $\mathbf{H}_{i,f}$ due to the fieldlike spin-pumping torques are defined in Eqs.~(3) and (4) of the main text.
By retaining only linear terms in the fluctuations of the magnetization around equilibrium, Eqs.~(\ref{S10}) and~(\ref{S11}) can be written in matrix form as
\begin{align}
Q\delta\mathbf{M}=0,
\end{align}
where $Q$ is an $8\times 8$ matrix and $\delta\mathbf{M}=(\delta m_A^x,\delta m_A^y,\delta m_B^x,\delta m_B^y,\delta m_C^x,\delta m_C^y,\delta m_D^x,\delta m_D^y)$. 
We solve the eight coupled equations numerically by requiring $\det(Q)=0$.

In the absence of spin pumping, the equilibrium orientations of the magnetizations are obtained by setting equilibrium conditions to Eqs.~(1) and~(2) of the main text, which is equivalent to minimizing the system's energy, i.e.,
\begin{align}
    E=\sum_{i=A,B,C,D} \mathbf{m}_{i} \cdot \mathbf{H}_{i,\text{eff}}\,,
    \label{s12}
\end{align}
with respect to the magnetic orientation of each layer. The effective field $\mathbf{H}_{i,\text{eff}}$ introduced in Eq.~(\ref{s12}) read as
\begin{align}
    \mathbf{H}_{A(D),\text{eff}}&= H_{0} \hat{y} - H_{E} \mathbf{m}_{B(C)} - M_{s} \left( \mathbf{m}_{A(D)} \cdot \hat{z} \right) \hat{z}\,, \label{HAD} \\
   \mathbf{H}_{B(C),\text{eff}}&= H_0\hat{y} - H_E\left(\mathbf{m}_{A(B)} + \mathbf{m}_{C(D)}\right) -M_s(\mathbf{m}_{B(C)} \cdot \hat{z})\hat{z}\,. \label{HCB} 
\end{align}
Due to the symmetry of the tetralayer samples, in static equilibrium, one finds $\mathbf{H}_{A(B),\text{eff}}=\mathcal{C}_{2y} \mathbf{H}_{D(C),\text{eff}}$, with $|H_{A(B),\text{eff}}| = |H_{D(C),\text{eff}}|$. Thus, it suffices to discuss the static equilibrium angles $\phi_A$ and $\phi_B$ as a function of exchange field ($H_E$) and external field ($H_0$). Figure S.1(a) shows that $\phi_A$ and  $\phi_B$ evolve in a non-linear manner with respect to external field and converge to $\pi$/2 rad upon saturation. The net fields~(\ref{HAD}) and~(\ref{HCB}) experienced by $\hat m_A$ and $\hat m_B$ in static equilibrium configuration as a function of external field for a constant value of interlayer exchange field, $\mu_0$$H_E=$ 24.4 mT is graphically represented in Figure S.2 (b). In the limit of high field, the net fields increase linearly with external field. For finite spin pumping, we find the (dynamical) equilibrium angle by including the fields $\mathbf{H}_{i,f}$, i.e., Eqs.~(3) and (4) of the main text, in the definition of $\mathbf{H}_{i,\text{eff}}$. 

\subsection{Distinguishing acoustic magnons from optical magnons}
In previous work, we examined the tetralayer model that is described above in the absence of interlayer spin pumping.  When interlayer spin pumping is not included in the equations of motion, there is a straightforward way to separate the optical magnons from acoustic magnons.  Indeed, this approach of labeling separately the acoustic and optical magnon branches is shown in Figure 3 within the main text.  We summarize this simplification of the tetralayer model now.

If we consider Eq.~(S.10), we can see that the linearized equation of motion for Layer B depends upon the dynamic amplitudes $\delta \mathbf{m}_{A}$, and $\delta \mathbf{m}_{C}$.  In the absence of spin pumping, we can exploit the symmetry of the tetralayer to only consider solutions where $\delta \mathbf{m}_C = \pm\delta \mathbf{m}_B$.  This substitution leads to a simplification of the problem as now only the determinant of a 4 $\times$ 4 matrix, which involves only Layers A and B, is needed for the magnon frequencies to be obtained.   If $\delta \mathbf{m}_C = +\delta \mathbf{m}_B$, then the acoustic magnon eigenfrequencies will be obtained.  Similarly, if $\delta \mathbf{m}_C = -\delta \mathbf{m}_B$, then the optical magnon eigenfrequencies will be obtained.

\section{Additional Magnetometry data}

Our synthetic antiferromagnets use permalloy as a ferromagnetic layer and ruthenium as the non-magnetic spacer material. Various samples (bilayers as well as tetralayers) are prepared by varying the spacer material thickness between $0.25$ nm to 3 nm in the interval of $0.25$ nm; the permalloy thickness is fixed to 5 nm. The samples are prepared using platinum both as seeding and capping layer or ruthenium as seeding layer and platinum as capping layer.  The magnetometry results are independent of the seeding material used, i.e., we observed no difference between using Pt or Ru as the seeding layer.  Since the RKKY interaction is oscillatory in nature, our samples are ferromagnetic for some spacer thicknesses, and antiferromagnetic for other thicknesses. The evolution of magnetization with external field for some typical ruthenium thicknesses for  tetralayer samples is shown respectively in Figure S.4.  The diamagnetic effect of the substrate has been removed in all the magnetometry results shown here.  The presence of hysteresis is a clear indication of ferromagnetic coupling, whereas absence of hysteresis, and zero magnetization at zero field, are indicative of antiferromagnetic coupling.


\section{Determining where the magnons spatially reside using micromagnetic simulations}

In this section, we use Mumax3\cite{vansteenkiste2014design} micromagnetic simulations to help identify where the acoustic and optical magnon branches are localized within the tetralayer, e.g. the surface or interior layers.  Extended details of our simulation methodology were reported elsewhere, when we computationally studied the acoustic and optical magnons of CrCl$_3$\cite{sklenar2021self}.  Here, we follow the same approach as our earlier work.  For these simulations, we have simply changed the material parameters to better model a synthetic antiferromagnet.

For computational ease we used a nanomagnet in our simulation.  The geometry we chose for our simulations is based on a standard example\cite{vansteenkiste2014design}, and is an ellipse of 160 nm long and 80 nm wide.  We use a saturation magnetization, $M_s$, of 700 $\times 10^3$ A/m, a ferromagnetic intralayer exchange stiffness of 13 $\times 10^{-12}$ J/m, and an antiferromagnetic interlayer exchange stiffness of -1.4 $\times 10^{-15}$ J/m.  The strength of the interlayer exchange stiffness was chosen to emulate the antiferromagnetic RKKY interaction in our materials.  In other words, these parameters reproduce the magnetization curves shown in Figure 1 (c) within the main manuscript.  

Before discussing our results we note that there are some quantitative discrepancies in the micromagnetic simulations compared with experiment.  This is due to the finite size effect of the nanomagnet we simulated, which is a proxy for a macroscopically large experimental sample.  We note that if we fine-tuned the interlayer exchange stiffness in the simulation, we would be able to better quantitatively match the experimental results.  \textit{We emphasize that the goal of these simulations is to qualitatively extract where the magnons spatially reside}.  Furthermore, one can compare these results to our earlier simulation results to see how these qualitative conclusions are  independent of the exact value of the interlayer exchange stiffness.

In Figure S.4 (a) and (b), we plot the numerically obtained acoustic magnon spectrum for the interior and surface layers, respectively.  In the vicinity of the avoided energy level crossing, it is clear that the low-energy acoustic magnon branch tends to reside on interior layers.  In contrast, the higher energy acoustic magnon branch tends to reside on the surface layers, especially in the vicinity of the avoided energy level crossing.  In Figure S.5 (a) and (b), we plot the optical magnon spectrum for the interior and surface layers, respectively.  In the vicinity of the avoided energy level crossing, it is clear that the low-energy optical magnon branch tends to reside on surface layers.  In contrast, the higher energy optical magnon branch tends to reside on the interior layers.

\section{Surface roughness characterization}
We characterized the surface roughness of both the bare substrate and the films that were deposited on top of the substrates for three samples.  Two representative images are shown in Figure S.6 (a) and (b) for the substrate and film surface roughnesses respectively.  Scans were taken over a $ 1 \times 1 $ $\mu m^2$ area.  The three average surface roughness values obtained for the substrate are 202, 178, and 229 pm.  The three average surface roughness measurements of the completed tetralayer are 147, 166, and 157 pm.  Thus, the surface roughness is reduced by 46 pm after the films are deposited, a $22\%$ reduction from the original roughness of the substrate.


\newpage

\begin{figure}
    \centering
    \includegraphics[scale=0.75]{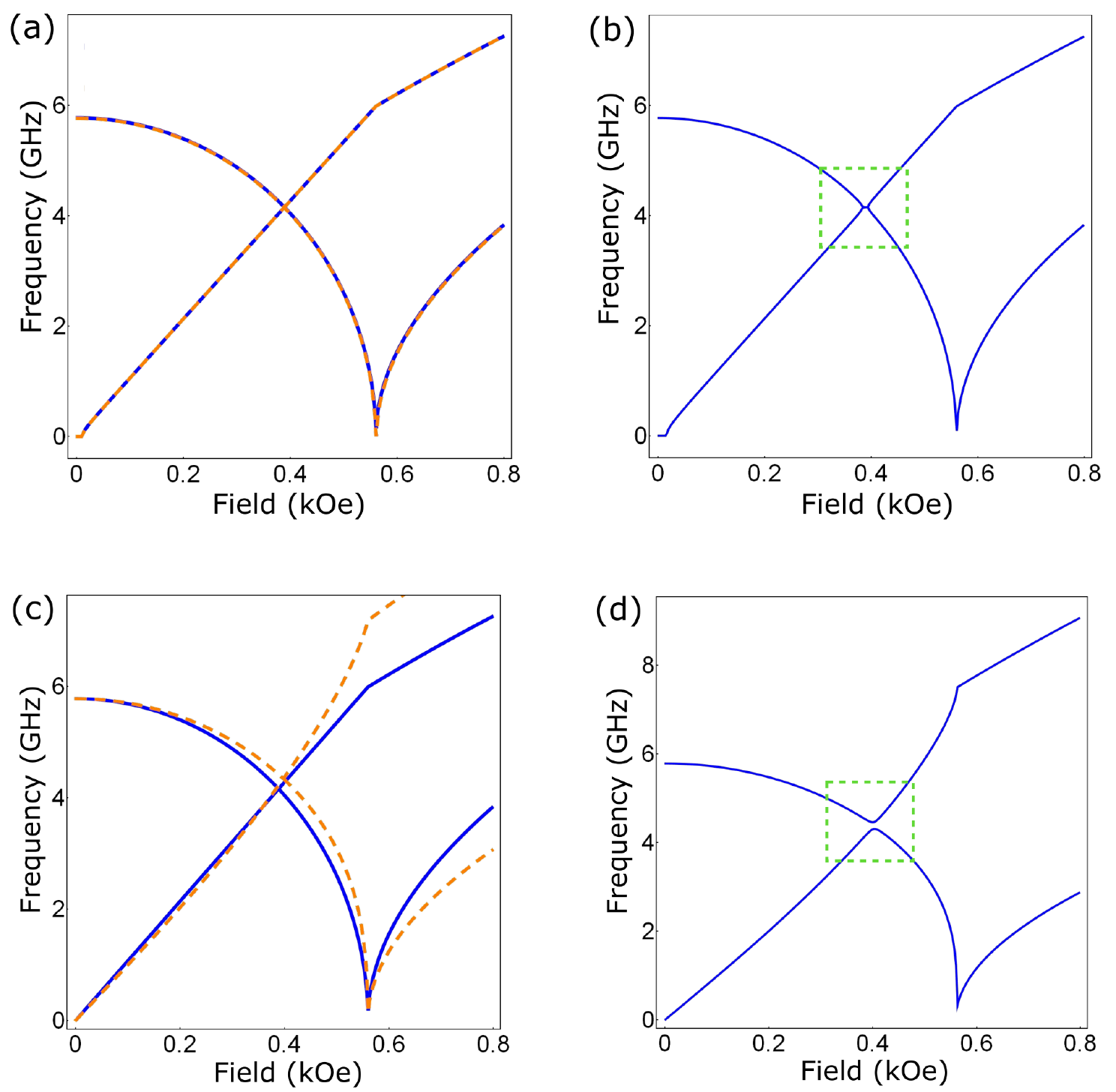}
    \caption{(a) Magnon spectrum of a symmetric bilayer for $\alpha_{A}=\alpha_{B}=0.01$ with  $\alpha_{d,AB}=0$ (blue solid line) and  $\alpha_{d,AB}=0.01$ (dashed orange line). (b) Magnon spectrum of an asymmetric bilayer with $\alpha_{A}=0.01$ and $\alpha_{B}=0.02$. The asymmetry of the local dissipation yields the level attraction highlighted in the dashed box.  In (c) and (d), we illustrate the role of the fieldlike spin pumping torques for a magnetic bilayer.  (c) The magnon bands (dashed orange line) of a symmetryic bilayer with $\alpha_{f,AB} = \beta_{f,AB} = 0.1$ are compared with the spectrum in the absence of dynamical torques (blue solid line).  (d) For an asymmetric bilayer, i.e., $\alpha_{f,AB} = 0.1$ and $\beta_{f,AB} = 0.15$, the level crossing between acoustic and optical bands is lifted.}
    \label{fig:my_label}
\end{figure}

\begin{figure}
    \centering
    \includegraphics[scale=0.35]{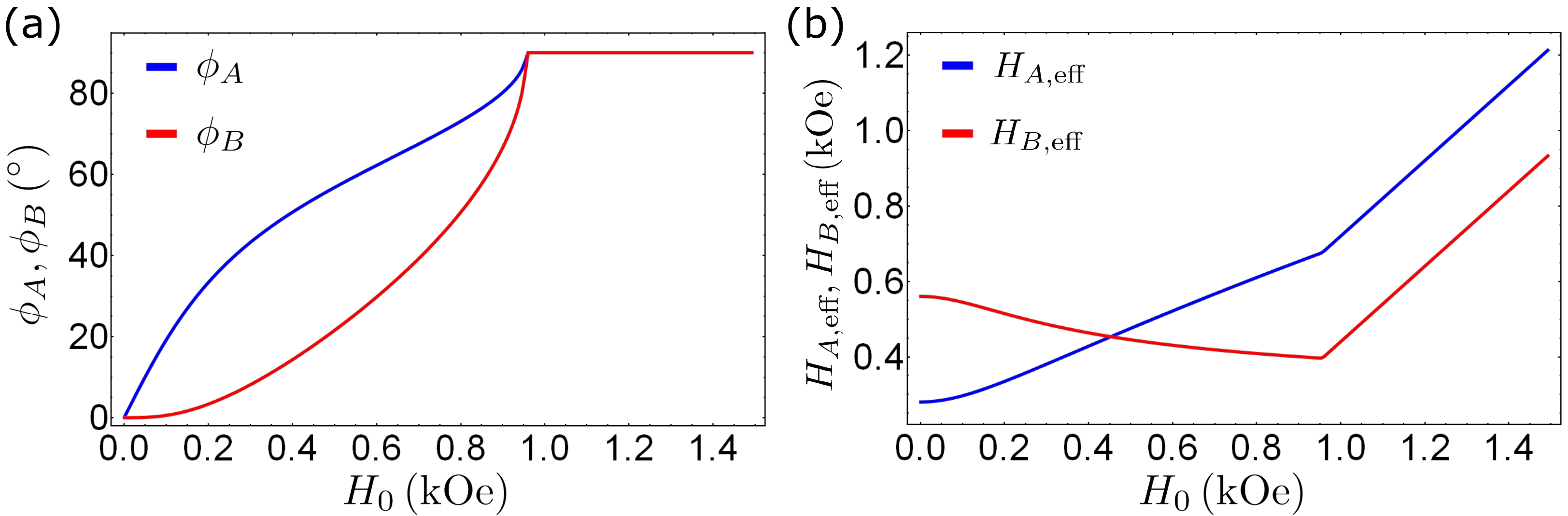}
    \caption{(a) Static equilibrium angles $\phi_A$ and  $\phi_B$ versus external field $H_0$. The angles vary non-linearly and converge to $\frac{\pi}{2}$ rad upon saturation. (b) Net field experienced by $\hat m_A$ and $\hat m_B$ at static equilibrium configuration i.e. $H_{A,eq}$, $H_{B,eq}$ versus external field $H_0$ for a constant value of interlayer exchange field: $\mu_0$$H_E=$ 24.4 mT. Initially, $H_{A,eq}$ decreases with field and $H_{B,eq}$ increases with field, but both increase linearly with external field for higher field values.}
    \label{fig:my_label}
\end{figure}


\begin{figure}
    \centering
    \includegraphics[scale=1]{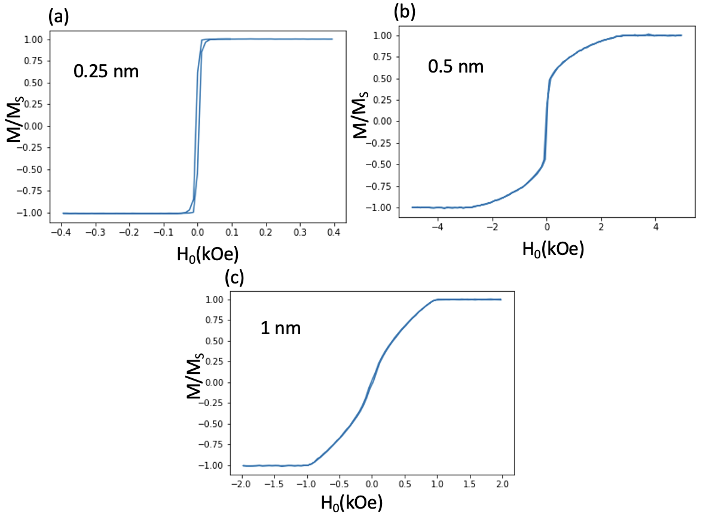}
    \caption{Normalized magnetization versus external field for tetralayer samples:(a) 0.25 nm spacer thickness, coupling is ferromagnetic; (b) 0.5 nm spacer thickness, coupling is ferromagnetic; (c) 1 nm spacer thickness, coupling is antiferromagnetic. The presence of hysteresis is a clear indication of ferromagnetic coupling, whereas absence of the same represents antiferromagnetic coupling. As it is
apparent in the graphs, coupling is antiferromagnetic at the spacer thickness of 1 nm. Moreover, the small loop indicating a residual ferromagnetism seen in asymmetrical bilayer samples has been removed here by using symmetrical samples with seeding and capping layers. }
\end{figure}


\begin{figure}
    \centering
    \includegraphics[scale=.2]{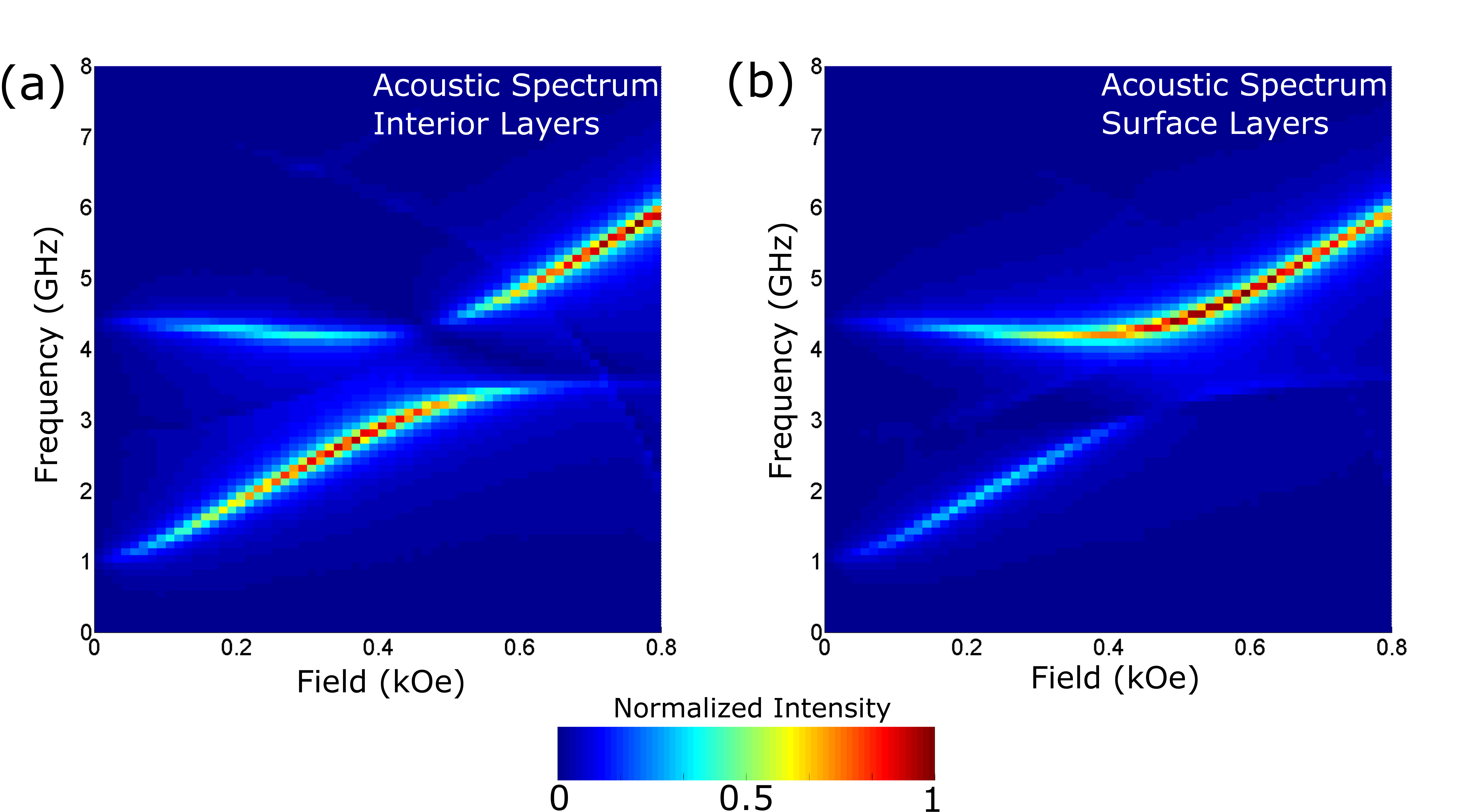}
    \caption{(a) The acoustic magnon energy spectrum of the tetralayer is plotted for the interior layers.  (b) The acoustic magnon energy spectrum of the tetralayer is plotted for the surface layers.  }
\end{figure}

\begin{figure}
    \centering
    \includegraphics[scale=.2]{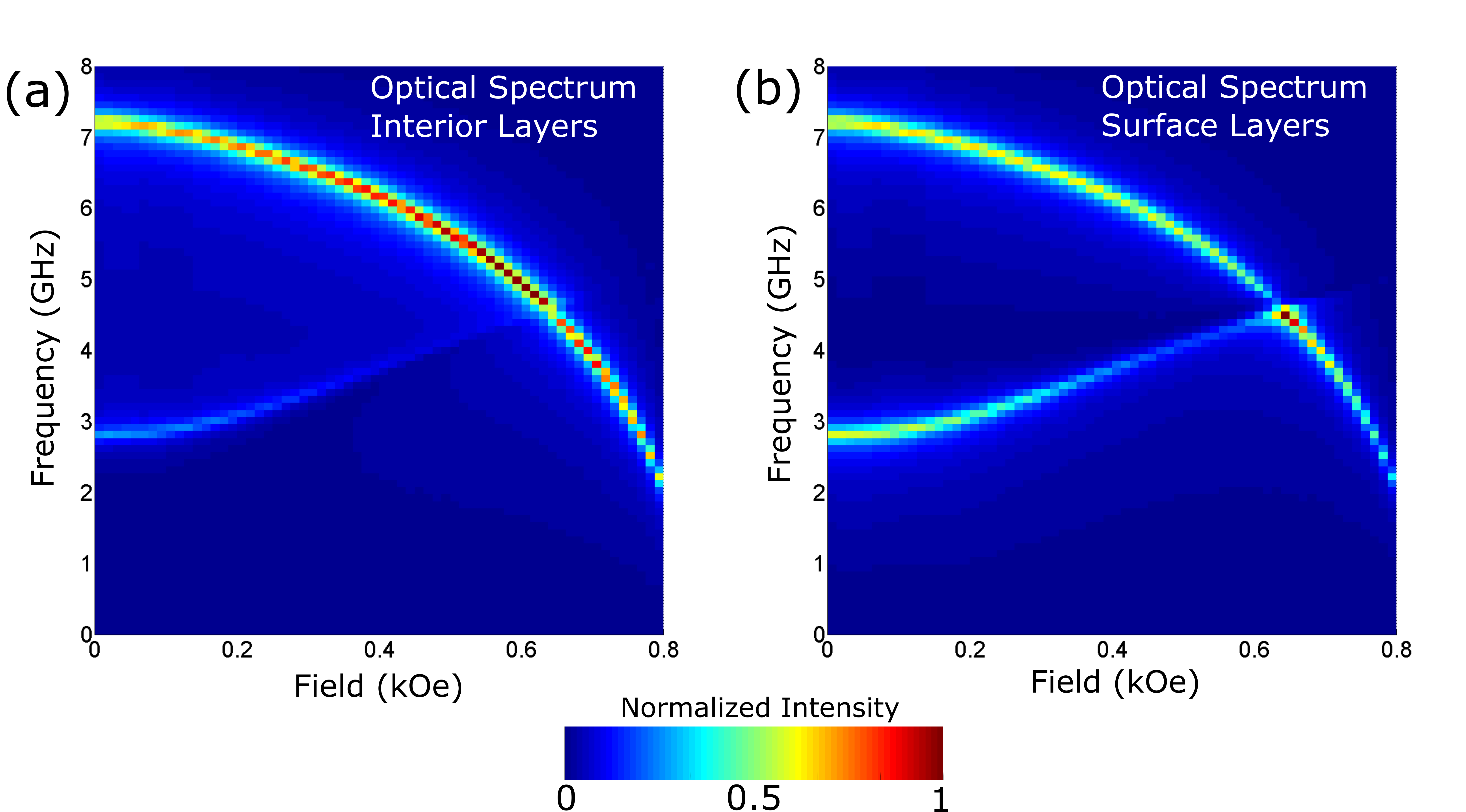}
    \caption{(a) The optical magnon energy spectrum of the tetralayer is plotted for the interior layers.  (b) The optical magnon energy spectrum of the tetralayer is plotted for the surface layers.  }
\end{figure}

\begin{figure}
    \centering
    \includegraphics[scale=.25]{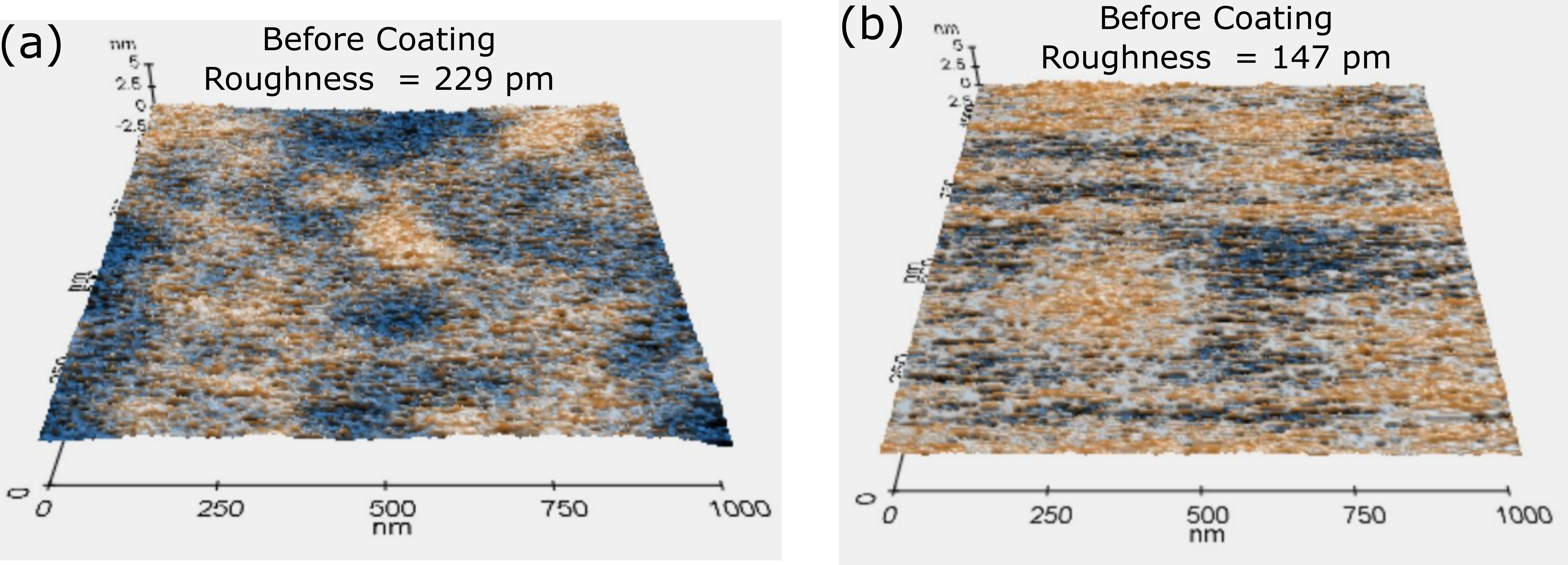}
    \caption{(a) and (b) show representative topographical images of the surface roughness of the sample substrate and the substrate after being coated with a tetralayer respectively.   }
\end{figure}